# Short-term forecasts of COVID-19 spread across Indian states until 1 May 2020

Neeraj Poonia[1], Sarita Azad[2*]

[1] School of Basic Sciences, Indian Institute of Technology Mandi, 175075, India.
[2] School of Basic Sciences, Indian Institute of Technology Mandi, 175075, India.

*Corresponding author sarita@iitmandi.ac.in

**Abstract**

The very first case of corona-virus illness was recorded on 30 January 2020, in India and the number of infected cases, including the death toll, continues to rise. In this paper, we present short-term forecasts of COVID-19 for 28 Indian states and five union territories using real-time data from 30 January to 21 April 2020. Applying Holt's second-order exponential smoothing method and autoregressive integrated moving average (ARIMA) model, we generate 10-day ahead forecasts of the likely number of infected cases and deaths in India for 22 April to 1 May 2020. Our results show that the number of cumulative cases in India will rise to 36335.63 [PI 95% (30884.56, 42918.87)], concurrently the number of deaths may increase to 1099.38 [PI 95% (959.77, 1553.76)] by 1 May 2020. Further, we have divided the country into severity zones based on the cumulative cases. According to this analysis, Maharashtra is likely to be the most affected states with around 9787.24 [PI 95% (6949.81, 13757.06)] cumulative cases by 1 May 2020. However, Kerala and Karnataka are likely to shift from the red zone (i.e. highly affected) to the lesser affected region. On the other hand, Gujarat and Madhya Pradesh will move to the red zone. These results mark the states where lockdown by 3 May 2020, can be loosened.

## 1 Introduction

COVID-19 illness, an on-going epidemic, started in Wuhan city, China, in December 2019 continues to cause infections in many countries around the world [1]. Considering the scale and speed of transmission of COVID-19, on 11 March 2020, the World Health Organization (WHO) declared it as a pandemic [2]. Thereafter, COVID-19 has become a threat to human life on the planet. It has shown rapid infections in almost all countries, and there is no cure available for this deadly virus. Presently governments have issued precautionary measures such as social distancing, sanitization of streets and markets, quarantine of suspected and infected cases, and lockdown of the communities at different scales (colonies, towns, states, and countries, etc.). In India, exponential growth has not been observed as compared to the USA and other European countries. It is due to the measures taken by the Indian government. It indicates that there is a strong influence of these measures, such as lockdown on the transmission behavior of COVID-19. On the other side, these measures create substantial economic losses to the communities, and hence actions mentioned above cannot be imposed for longer periods. Mainly, developing countries (such as India) cannot afford such payoff after some finite time. The Indian government has continuously reviewed every hour situation in every state. The government has become more focused on localizing the lockdown in particularly alarming states and few towns which are hotspots for COVID-19. For all these, it is important to have short-term forecasts which can be steering point for decision-makers and administrations. In this connection, data-based statistical models such as Autoregressive integrated moving average (ARIMA) and Holts method have shown effectiveness in predicting short-term forecast including the dengue fever [3, 4], the hemorrhagic fever with renal syndrome [5], Tuberculosis [6] and COVID-19 [7]. ARIMA has more ability compared to other prediction models like the support vector machine and wavelet neural network for drought forecasting [8]. Also, exponential smoothing methods have been widely used for forecasting of the population in West Java [9], an inflation rate of Zambia [10] including a prediction for epidemic mumps [11] and COVID-19 [12–17]. However, mainly, for India, the short-term forecast is not done thoroughly. As India has diversity across the states, it will be essential to study the spreading behavior of COVID-19 in different Indian states. This article presents a short-term forecast for various Indian states which are severely infected.

The main objective of the present paper is to present 10-day ahead forecasts from 22 April to 1 May 2020 of the cumulative number of infected cases and deaths due to COVID-19. This work also presents the analysis of Indian states at the regional level to understand the spread of infection. The current situation of India is shown in Figure 1, with the cumulative number of infected cases and deaths from 30 January to 21 April 2020.

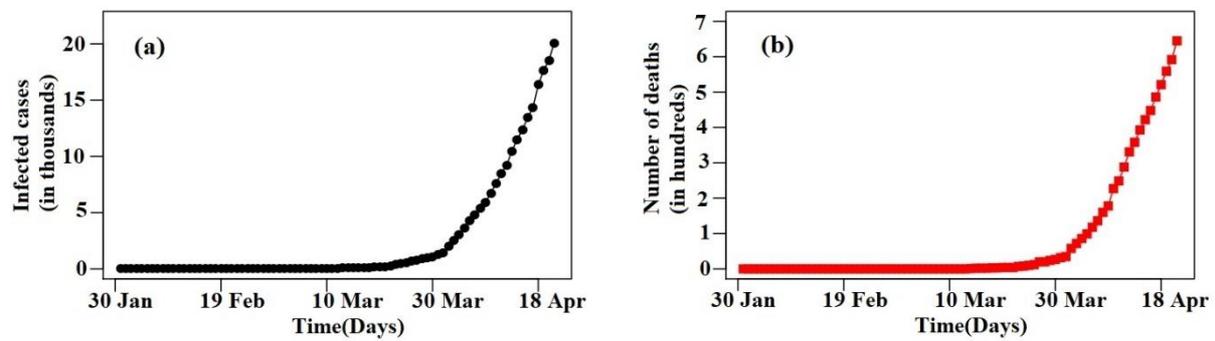

**Figure 1:** (a) Number of infected cases from 30 January to 21 April 2020; (b) Number of deaths from 30 January to 21 April 2020.

## 2 Materials and Methods

### 2.1 ARIMA Model

The process whose statistical properties do not change with time, i.e. process with constant mean and constant variance, known as a stationary process, is a crucial collection of stochastic processes. Mathematically, the joint distribution of $X(t_1),...,X(t_k)$ and $X(t_{1+\tau}),...,X(t_{k+\tau})$ is the same for all $t_1, t_2,..., t_k$ of a stationary process. Simply put, shifting the origin of time by a quantity $\tau$ does not change the statistical properties of the process. Usually, dealing with real-time data, most time series does not exhibit stationarity in nature as they have no fixed mean. The properties of the crucial collection of models for which the $d^{th}$ difference of the time series is a stationary mixed autoregressive moving average process (ARMA). These models are known as ARIMA models. The ARMA model, introduced by Box and Jenkins, is the collection of popular methods that are directly applicable to modeling and analyzing the time series [18]. The ARMA model is formed by the merger of two models, the autoregressive AR(p) model and the moving average MA(q) model. These models are directly applicable to time series with stationary behavior. In case the series is non-stationary, it must be dealt via differencing to make it stationary. Generally, the ARMA model after differencing is known as ARIMA (p, d, q). Addressing

$$H_t = \nabla^d X_t = (1-k)^d X_t \qquad (1)$$

The general ARIMA model is given by

$$H_t = \alpha_1 H_{t-1} + ... + \alpha_p H_{t-p} + J_t + ... + \beta_q J_{t-q} \qquad (2)$$

Hence, the ARIMA model can be written as

$$f(k)H_t = g(k)J_t \qquad (3)$$
$$f(k)\nabla^d X_t = g(k)J_t \qquad (4)$$

The expressions in the Eq. 4 are defined as: $f(k), g(k)$ are polynomials of degree $p$, $q$ respectively s.t.

$$f(k) = 1 - \alpha_1 k - ... - \alpha_p k^p \qquad (5)$$

and

$$g(k) = 1 + \beta_1 k + ... + \beta_q k^q \qquad (6)$$

While, $\nabla^d$ is an operator, known as difference operator, and used to make the difference of time series stationary; and $d$ is the difference value. In real-time data, taking the first difference ($d=1$) is usually found to be sufficient and occasionally second difference ($d=2$) would be enough to achieve stationarity.

Akaike Information Criterion (AIC) is one of the essential criteria to select between competing models. Mathematically,

$$\text{AIC} = \log\left(\frac{\sum_{t=1}^{T} e_t}{T}\right) + \frac{2p}{T} \qquad (7)$$

The model which has the least AIC is selected as the best model. Autocorrelation functions (ACF) and partial autocorrelation functions (PACF) are used to select order of moving average process MA(q) and autoregressive process AR(p) respectively. In the process to investigate the stationarity of time series Kwiatkowski–Phillips–Schmidt–Shin (KPSS) [19] and Augmented Dickey-Fuller (ADF) [20] tests are used. To reject the null hypothesis, the *p*-value must be smaller than the significance level.

### 2.2 Holt'sMethod

The numbers of confirmed cases and deaths in India are increasing day by day, as shown in Figure 1 thereupon the time series exhibit trend. Simple exponential smoothing methods should not apply in this case. When data shows the pattern, and there is no seasonality, Holt's method is a primary tool to handle it. Holt's method is a double exponential smoothing method (not based on ARIMA approach) which has two parameters. This method divides the time series into two sections: the level and the trend denoted by $B_t$ and $M_t$ respectively. These two parts are as follows:

$$B_t = \alpha X_t + (1-\alpha)(B_{t-1} + M_{t-1}) \qquad (8)$$
$$M_t = \gamma(B_t + B_{t-1}) + (1-\gamma)M_{t-1} \qquad (9)$$

The in-future forecasts values $X_{t+h}$ of the time series can be calculated by:

$$X_{t+h} = B_t + M_t(h) \qquad (10)$$

where *h* is the number of periods in the future. Diverse statistical meaning-making models in the R-language platform were used to evaluate the time series of infected cases and deaths for prediction purposes.

3. Results and Discussion

We present results for 10-day ahead forecasts (22 April to 1 May2020) generated for the cumulative number of infected cases and deaths in India as well as in the ten most affected states: Kerala, Maharashtra, Delhi, Gujarat, Tamil Nadu, Telangana, Uttar Pradesh, Madhya Pradesh, Karnataka, Rajasthan. In this work, we used two models Holt's method and ARIMA model to forecast the cumulative infected cases and deaths of COVID-19. For the ARIMA model, we forecast per day new infected case(s) and new death(s), whereas for Holt's method cumulative numbers are generated.

**3.1 Validation:** As of 12 April 2020, there were 9205 cumulative numbers of infected cases and 331 cumulative numbers of deaths in India. For validation purposes, we forecasted the cumulative number of infected cases and deaths from 7-12 April 2020, using ARIMA and Holt's method. Our forecasting results showed 8064 [PI 95% (6743, 9842)] cumulative number of infected cases using ARIMA(1,1,0) model and 8936 [PI 95% (7424, 10588)] cases using Holt's method ($\alpha=0.9464$, $\beta=0.4311$), in both cases the 95% prediction intervals includes the actual values. While forecasting results of the cumulative number of deaths are 287 [PI 95% (210, 377)] using Holt's method ($\alpha=0.8052$, $\beta=0.2905$) which includes the actual value within 95% prediction interval but ARIMA(0,1,1) underfit the data with 276 [PI 95% (225, 326)] cumulative number of deaths. To stabilize the variance in Holt's and ARIMA(1,1,0) square root transformation is used.

**3.2 India forecasting:**
**3.2.1 ARIMA model:** During the analysis and forecasting of a time series, it is good to plot the time series data and pay attention to the unique features exhibited by the time series. It gives direction to the researcher for choosing an appropriate modeling approach that directly captures identified features. Before starting the procedure, there is a need to make the time series stationary. To stabilize the variance, we used square root transformation on the infected number of cases per day time series. For investigating the stationarity of time series, we take the support of the KPSS and ADF test, and results are shown in Table 1. The first difference of series, i.e. $d=1$, is optimum to make series reasonably stationary. Based on a 5% significance level both the tests, ADF and KPSS, reject the hypothesis of stationarity of time series without making any difference. Afterwards taking the first difference, both the criteria agree on the stationarity of time series. Further, to estimate another two parameters of the candidate model, the ACF and PACF of series, first difference, and square root transformation are used. From Figure 2(a) and 2(b), the ACF display one spike, and the PACF also displays one spike. Initially, on the bases of the number of spikes, we selected ARIMA(1, 1, 1). Alternate models are also used to compete with the ARIMA(1,1,1) model. All alternative models and their AIC values with the Ljung-Box test *p*-values are shown in Table 2. A model with a minimal amount of AIC is to have well-behaved residuals. Finally, we select ARIMA(1,1,2) for forecasting. In terms of the residuals, the ARIMA(1,1,2) model passed the Ljung-Box test with *p*-values larger than 0.05 level of significance. Since ARIMA(1,1,2) has the lowest AIC value, which means the residuals of ARIMA(1,1,2) are much well behaved compared to other considered models. We examine that all the residuals are scattered around zero mean with constant variance. Using this, ARIMA(1,1,2) model observe 36335.53 [95% PI(30884.56 -42918.87)] cumulative infected cases between by 1 May2020, results are shown in Table 3.

**Table 1**: Table of *p*-values from ADF and KPSS tests after taking the differences of square root transformed data for infected cases per day in India.

| Number of Difference | ADF test (*p*-value) | KPSS test (*p*-value) |
|---|---|---|
| *d*=0 | 0.961 | 0.01 |
| $d=1$ | 0.01 | 0.058 |

**Table 2:** Potential models for infected cases per day with AIC value and Ljung-Box test *p*-value.

| Model | AIC value | Ljung-Box test (*p*-value) |
|---|---|---|
| ARIMA (0,1,2) | 427.77 | 0.263 |
| ARIMA (1,1,2) | 418.82 | 0.518 |
| ARIMA (0,1,1) | 428.05 | 0.341 |
| ARIMA (1,1,1) | 428.59 | 0.258 |
| ARIMA (1,1,3) | 420.72 | 0.438 |

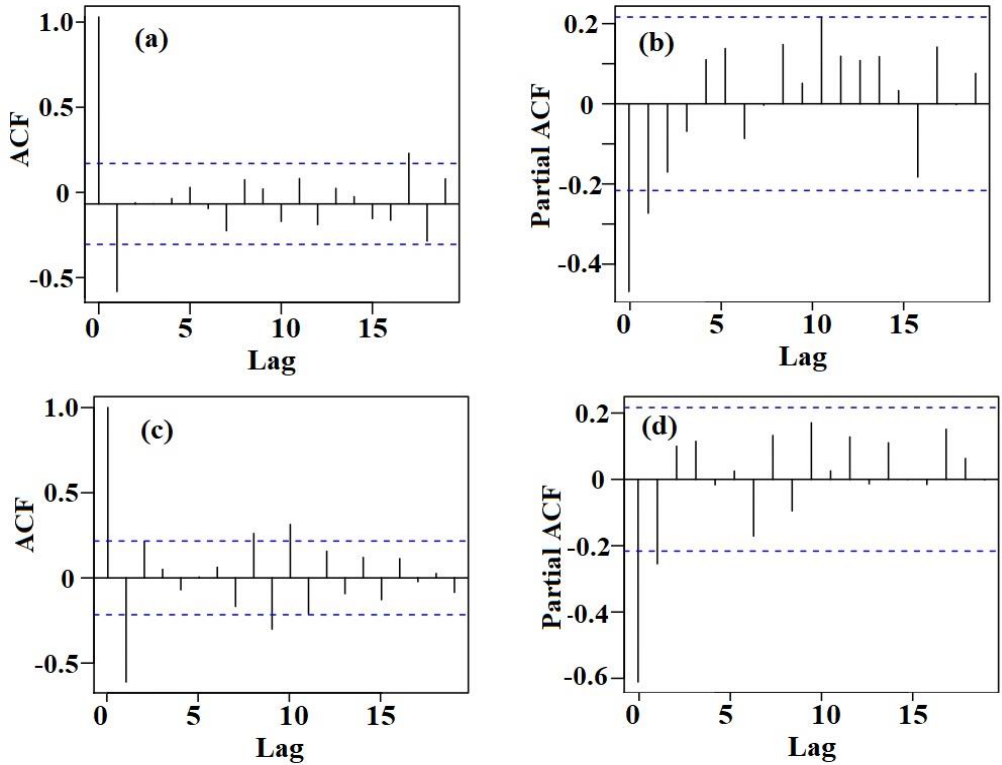

**Figure 2:** (a) ACF for the infected number of cases per day after square root transformation; (b) PACF for the infected number of cases per day after square root transformation; (c) ACF for the number of deaths per day; (d) PACF for the number of deaths per day.

**Table 3:** Results of 10-days ahead forecasts (22 April to 1 May 2020) using ARIMA model for the cumulative number of infected cases and deaths.

| Date | Forecast of cumulative cases | 95% PI for infected cases | Forecast of deaths per day | 95% PI for deaths |
|---|---|---|---|---|
| 22 April 2020 | 21507.35 | (21099.49, 21983.68) | 680.46 | (671.04, 1004.66) |
| 23 April 2020 | 22980.52 | (22145.61, 23956.81) | 727.64 | (708.79, 1061.28) |
| 24 April 2020 | 24498.87 | (23213.42, 26004.80) | 772.26 | (741.49, 1117.82) |
| 25 April 2020 | 26061.76 | (24297.82, 28133.38) | 817.41 | (773.69, 1175.92) |
| 26 April 2020 | 27668.50 | (25393.68, 30348.45) | 863.09 | (805.46, 1235.49) |
| 27 April 2020 | 29318.42 | (26496.10, 32655.88) | 909.29 | (836.87, 1296.49) |
| 28 April 2020 | 31010.80 | (27600.44, 35061.35) | 956.02 | (867.97, 1358.85) |

| 29 April 2020 | 32744.95 | (28702.47, 37570.28) | 1003.28 | (898.79, 1422.54) |
| 30 April 2020 | 34520.13 | (29798.33, 40187.84) | 1051.06 | (929.39, 1487.52) |
| 1 May 2020 | 36335.63 | (30884.56, 42918.87) | 1099.38 | (959.77, 1553.76) |

Since only one difference makes the time series stationary, we conclude to take *d*=1. Results of ADF and KPSS tests are presented in Table 4. From Figure 2(c) and 2(d), ACF demonstrates two significant spikes, and PACF demonstrates zero significant spike. Based on the number of spikes, we selected ARIMA (0, 1, 2). Alternate models were also used to compete with the ARIMA (0,1,2) model. Details of other potential models along with AIC values and Ljung-Box test *p*-values given in Table 5. Furthermore, to forecast the number of deaths per day in India, we found ARIMA (0,1,3) a reasonable model among other competitor models it has minimum AIC value. Furthermore, we found residuals are randomly scattered around zero mean with non-changing variance with time. Also, ARIMA(0,1,3) does not show a lack of fit with the Ljung-box test *p*-value larger than 0.05. Graphical results of forecasting from infected cases and deaths are shown in Figure 3. Applying ARIMA(0,1,23), 1099.38 [95% PI(959.77-1553.76)] cumulative deaths are expected in coming 10 days in India. Results for 10-day ahead forecast for per day infected cases, and deaths are shown in Table 3. To eliminate the effect of square root transformation in per day infected cases we take a square of forecasted observations.

**Table 4**: Table of *p*-values from ADF and KPSS tests for deaths per day in India.

| Number of Difference | ADF test (*p*-value) | KPSS test (*p*-value) |
|---|---|---|
| *d=0* | 0.979 | 0.01 |
| *d=1* | 0.01 | 0.058 |

**Table 5:** Potential models for deaths per day data with AIC values and Ljung-Box test *p*-values.

| Model | AIC value | Ljung-Box test (*p*-value) |
|---|---|---|
| ARIMA (0,1,3) | 497.32 | 0.408 |
| ARIMA (1,1,4) | 499.79 | 0.208 |
| ARIMA (0,1,2) | 498.10 | 0.248 |
| ARIMA (1,1,2) | 498.77 | 0.274 |
| ARIMA (1,1,3) | 498.41 | 0.365 |

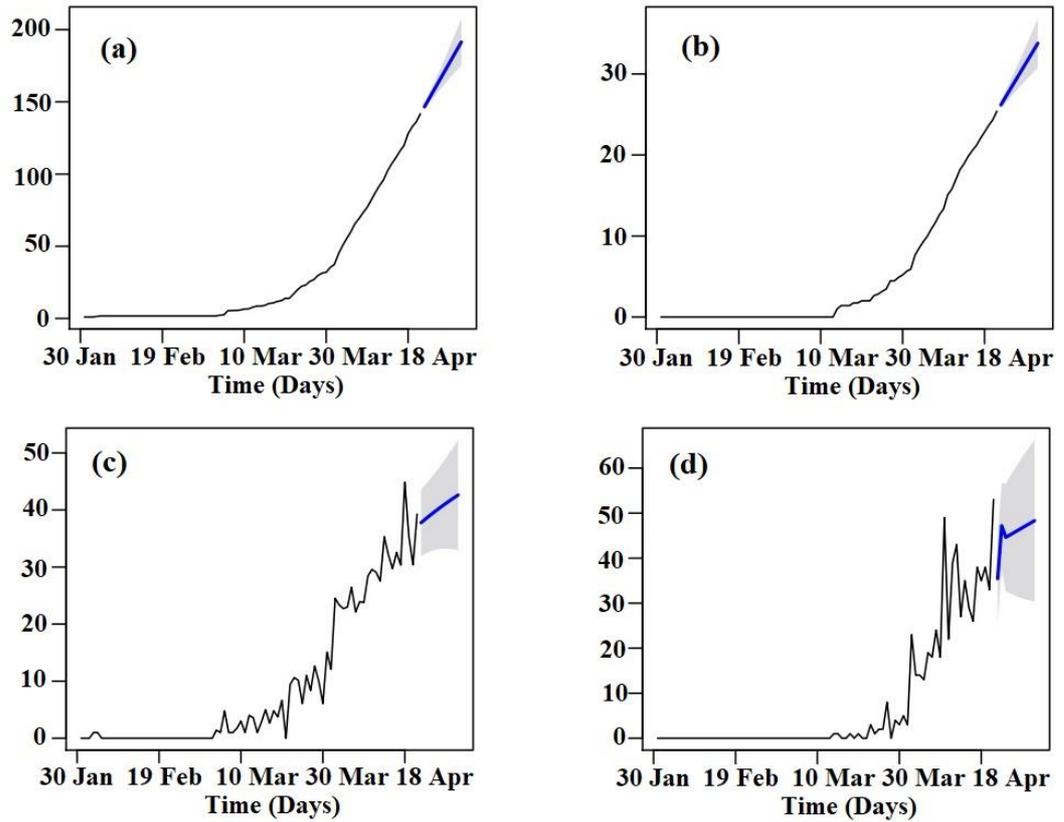

**Figure 3:** (a)10-days ahead forecast (22 April to 1 May 2020) for the number of infected cases per day using ARIMA(1,1,2) model; (b) 10-days ahead forecast (22 April to 1 May 2020) for the number of deaths per day using ARIMA(0,1,3) model; (c) 10-days ahead forecast (22 April to 1 May 2020) using Holt's method for the cumulative number of infected cases; (d) 10-days ahead forecast (22 April to 1 May 2020) for the cumulative number of days using Holt's method.

**3.2.2 Holt's Method:** The time series plot of the cumulative number of confirmed cases and deaths for India is presented in Figure 1 exhibiting the trend in time series, but it does not have a pattern of seasonality. As a result of the features shown by time series in Figure 1, Holt's method was selected in this study to accomplish a 10-day ahead forecast (22 April to 1 May 2020). Generally, a Holt method has two smoothing constants, $\alpha$, and $\beta$ (their values lie in range 0 and 1). The square root transformation is used to stabilize the variance in the time series of infected cases. In the process to attain the optimal parameters we applied by trial and error technique. Results are shown in Table 6 with the value of $\alpha$, $\beta$, AIC, and RMSE values. The best model is selected with the lowest AIC and RMSE values. With the parameters, $\alpha=0.9$ and $\beta=0.3$, obtained values of AIC and RMSE are 381.02 and 1.05, respectively. For this model, Ljung-Box test $p$-value=0.468 which agrees that model does not exhibit any lack of fit.

Using Holt's method, different values of $\alpha$ and $\beta$ are tried to retrieve the optimum forecast for cumulative deaths. The square root transformation is used to stabilize the variance in the time series of deaths. The results of the trials are listed in Table 7 with AIC and RMSE values. Smallest values of AIC=151.78 and RMSE=0.26 at $\alpha=0.8$ and $\beta=0.2$ are achieved. Subsequently, checking the Ljung-Box test $p$-value=0.109 we identify that model does not lack of fit. Graphical results of forecasting from infected cases and deaths are presented in Figure 3. From Table 8, 36624.43 [95% PI(30716.59-43051.56)] cumulative infected cases and 1140.70 [ PI % (945.32-1354.42)] cumulative deaths are in India up-to 1 May2020.

**Table 6:** Selection process for parameters in Holt's method to forecast the cumulative number of infected cases in India.

| α | β | AIC value | RMSE |
|---|---|---|---|
| 0.1 | 0.1 | 503.15 | 2.19 |
| 0.5 | 0.1 | 435.19 | 1.46 |
| 0.5 | 0.5 | 400.55 | 1.18 |
| 0.9 | 0.5 | 383.37 | 1.06 |
| 0.9 | 0.3 | 381.02 | 1.05 |

**Table 7:** Selection process for parameters in Holt's method to forecast the cumulative number of deaths.

| α | β | AIC value | RMSE |
|---|---|---|---|
| 0.1 | 0.1 | 253.11 | 0.49 |
| 0.5 | 0.1 | 185.53 | 0.32 |
| 0.5 | 0.5 | 167.63 | 0.29 |
| 0.9 | 0.5 | 157.12 | 0.27 |
| 0.8 | 0.2 | 151.78 | 0.26 |

**Table 8:** Results of 10-days ahead forecasts (22 April to 1 May 2020) using Holt's method for the cumulative number of infected cases and deaths.

| Date | Forecast of cumulative infected cases | 95% PI for infected cases | Forecast of cumulative deaths | 95% PI for deaths |
|---|---|---|---|---|
| 22 April 2020 | 21498.39 | (20883.67, 22122.02) | 685.95 | (658.43, 714.02) |
| 23 April 2020 | 22981.26 | (21992.15, 23992.12) | 730.79 | (690.78, 771.94) |
| 24 April 2020 | 24513.58 | (23102.62, 25966.35) | 777.06 | (723.16, 832.91) |
| 25 April 2020 | 26095.35 | (24209.69, 28051.72) | 824.75 | (755.44, 897.10) |
| 26 April 2020 | 27726.57 | (25311.47, 30251.70) | 873.86 | (787.58, 964.62) |
| 27 April 2020 | 29407.24 | (26407.03, 32568.84) | 924.39 | (819.55, 1035.54) |
| 28 April 2020 | 31137.36 | (27495.78, 35005.34) | 976.34 | (851.32, 1102.92) |
| 29 April 2020 | 32916.93 | (28577.24, 37563.27) | 1029.71 | (882.88, 1187.82) |
| 30 April 2020 | 34745.95 | (29650.98, 40244.67) | 1084.49 | (914.22, 1269.30) |
| 1 May 2020 | 36624.43 | (30716.59, 43051.56) | 1140.70 | (945.32, 1354.42) |

### 3.3 Indian states forecasting:

COVID-19 is spreading very fast in India. Locating the regions of most spread within India will give insight for the lifting the lockdown which commenced on 25 March 2020. On the regional level, this study shows the analysis for the cumulative number of cases but not deaths due to the unavailability of data. A glimpse of the current situation of the increasing number of cases in 10 states is given in Figure 4, certainly detectable that Maharashtra, Gujarat, and Delhi are the most affected states in India till 21April2020. And Kerala is least affected in our list of states. Time series starts from the date when the first case was reported in the respective state.

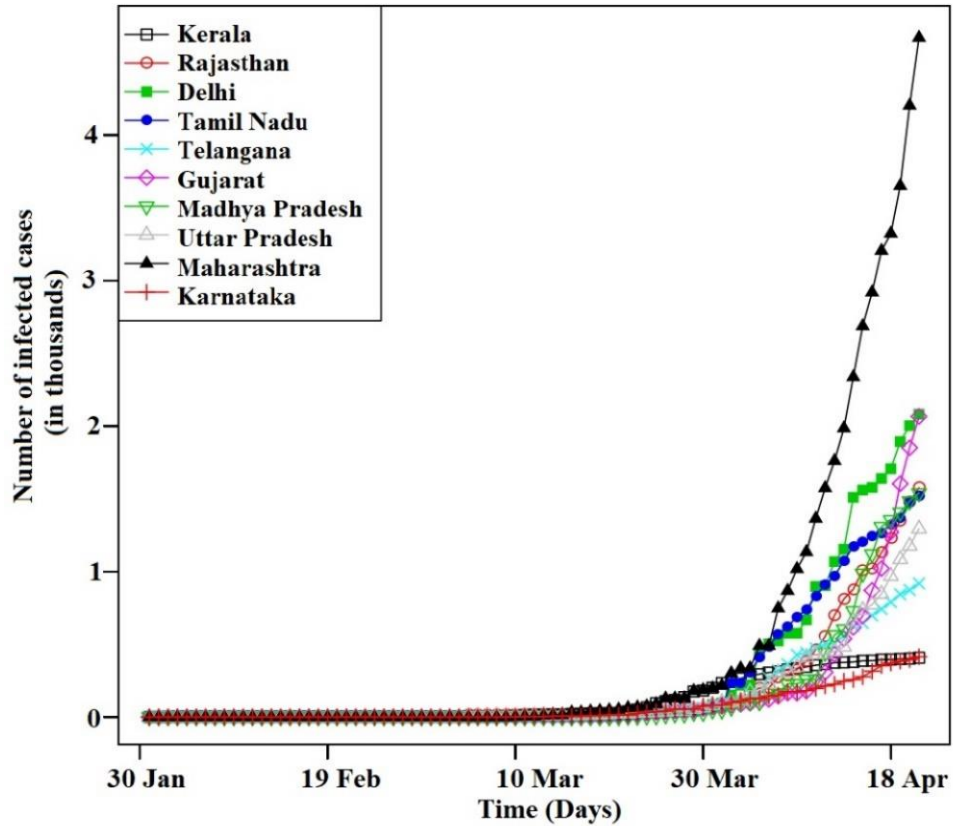

**Figure 4:** Number of infections in the ten most affected Indian states by corona-virus as of 30 January to 21April2020.

**3.3.1 ARIMA model:** For forecasting purposes, using the ARIMA model, the number of newly infected cases per day are analyzed instead of cumulative infected cases. To select the optimum ARIMA model for each state, firstly each state's time series is made stationary by taking differences. Next, we used ADF and KPSS tests to check stationarity. To stabilize the variance of Delhi, Telangana, Uttar Pradesh, and Gujarat time series, cube root transformations are used; later, one difference is enough to remove the trend. While to stabilize the variance of Maharashtra, Karnataka and Rajasthan time series, square root and square transformations are used, respectively.The same procedure is adopted for all the ten-time series of infected cases per day. AIC values are used to select the best models, and the model is chosen on the base of the smallest AIC value. Results of analysis for ARIMA models are shown in Table 9. Analysis by ARIMA models shows that Maharashtra and Gujarat will be the most affected states by 1 May2020, with around 9787.24 and 4216cumulative cases, respectively. As we observe that Kerala's growth is declining and it will be less affected states with 449 [PI 95%(408-574.99)] cumulative cases. All the models passed the Ljung-Box test as well as does not show any lack of fit.

**Table 9:** Region-wise details of ARIMA models which were used for 10-days ahead forecasts (22 April to 1 May 2020), along with AIC values and Ljung-Box test *p*-values. Point forecasts and 95% prediction intervals are given in the last two columns.

| Region | ARIMA Model | AIC value | Ljung-Box test (*p*-value) | Point forecast for infected cases | 95% PI for infected cases |
|---|---|---|---|---|---|
| Kerala | (2,1,0) | 498.53 | 0.329 | 449.38 | (408, 574.90) |
| Maharashtra | (0,1,2) | 233.65 | 0.807 | 9787.24 | (6949.81, 13757.06) |
| Rajasthan | (0,1,1) | 947.38 | 0.147 | 2741.40 | (2305.22, 3053.91) |
| Delhi | (1,1,2) | 177.46 | 0.064 | 3039.73 | (2139.72, 6085.18) |
| Telangana | (2,1,0) | 133.99 | 0.112 | 1321.37 | (940.84, 2740.89) |
| Karnataka | (3,1,0) | 160.03 | 0.371 | 565.74 | (419.09, 945.45) |
| Gujarat | (0,1,0) | 89.76 | 0.131 | 4216.00 | (2216.24, 13118.90) |
| Uttar Pradesh | (2,1,1) | 140.25 | 0.161 | 2652.21 | (1612. 43,4891.99) |
| Tamil Nadu | (1,1,1) | 440.69 | 0.840 | 2157.35 | (1520, 2878.82) |
| Madhya Pradesh | (0,1,1) | 340.62 | 0.961 | 2281.84 | (1540, 3688.99) |

**3.3.2 Holt's method:** Square root transformation is used to stabilize the variance of Rajasthan, Maharashtra, Karnataka,and Uttar Pradesh. The cube root and square transformation are used for Delhi, Kerala, Telangana, and Gujarat, respectively. Summary of Holt's method display that Maharashtra and Delhi will be most affected states with around 9768.91 and 3768.39cumulative number of infected cases, respectively. Meanwhile, Kerala will be the less affected state in our list with about 451.67 cumulative number of infected cases. The selection of optimum Holt's method is performed using the minimum values of AIC and RMSE. Although, all the model passed the Ljung-Box test, which state that model does not show any lack of fit. Results of the forecast for each state are given in Table 10 with Ljung-Box test *p*-values. The final graphical results of the analysis using both the models, ARIMA model, and Holt's method, are shown in Figures 5-11.

**Table 10:** Region-wise 10-days ahead forecasts (22 April to 1 May 2020) details of Holt's method, along with Ljung-Box test *p*-values. Point forecast and 95% prediction intervals are given in the last two columns.

| Region | Ljung-Box test (*p*-value) | Point forecast for infected cases | 95% PI for infected cases |
|---|---|---|---|
| Kerala | 0.134 | 451.67 | (408, 858.58) |
| Maharashtra | 0.776 | 9768.91 | (7453.81, 12396.63) |
| Rajasthan | 0.073 | 2978.53 | (1921.79, 4265.86) |
| Delhi | 0.051 | 3768.39 | (2081, 6607) |
| Telangana | 0.029 | 1424.42 | (919, 3171.29) |
| Karnataka | 0.166 | 602.05 | (495.65, 708.44) |
| Gujarat | 0.229 | 3562.28 | (2992.38, 4052.81) |
| Uttar Pradesh | 0.138 | 2569.51 | (1773.49, 3512.69) |
| Tamil Nadu | 0.635 | 2158.51 | (1664.95, 2652.07) |
| Madhya Pradesh | 0.162 | 2301.68 | (1540, 3321.74) |

**3.4 Recommendations on Lockdown Extension:** India comprises 28 states and eight union territories. Here we have analyzed all the states, including five union territories. In Figure 12, the spatial distribution of coronavirus outbreak shows eight states in the red zone (extremely affected), namely, Delhi, Rajasthan, Uttar Pradesh, Maharashtra, Telangana, Karnataka, Kerala, Tamil Nadu.
Similarly, seven states in the blue zone (intermediate affected), are Jammu & Kashmir, Punjab, Haryana, Gujarat, Madhya Pradesh, Andhra Pradesh, West Benga. The green and light green (least affected) zones include Himachal Pradesh, Uttrakhand, Bihar, Jharkhand, Chhattisgarh, Odisha, Sikkim, Arunachal Pradesh, Assam, Nagaland, Manipur, Mizoram, Tripura, Meghalaya, Goa. To construct the zones, we have divided the cumulative cases of states into quartiles as on 1 April 2020. The same procedure is carried out for forecasted cumulative cases until 1 May 2020.As infected cases are increasing, it is essential to notice which of the states will shift their zone.

Figure 13 shares Delhi, Rajasthan, Uttar Pradesh, Gujarat, Madhya Pradesh, Maharashtra, Telangana, Tamil Nadu in the red zone and Jammu & Kashmir, Punjab, Haryana, Kerala, Karnataka, West Bengal in the blue zone while Himachal Pradesh, Goa, Uttrakhand, Bihar, Jharkhand, Chhattisgarh, Odisha, Sikkim, Assam, Arunachal Pradesh, Nagaland, Manipur, Mizoram, Tripura, Meghalaya are in green and light green zones.

It is found that Kerala and Karnataka were in the red zone, and Gujarat and Madhya Pradesh were in the blue area until 1 April 2020 (Figure 12). But they are likely to change their positioning by 1 May. Accordingly, Kerala and Karnataka will shift to the blue zone as cases are declining in both states. Conversely, Gujarat and Madhya Pradesh will move to the red area. The government should impose extra precautions in these states, as the cases will significantly rise in both in the coming days. While lockdown should remain in the red zone, conversely, the blue area is not remarkably affected by COVID-19, so lockdown should be lifted with some restrictions. It is advisable to lift the lockdown in states within green and light green zones for the proper functioning of the economy.

Further, analysis of red and blue zones at the regional level is of importance to decide about raising the district wise lockdown.

## 4 Conclusions

The spread of the COVID-19 epidemic has been slow in India as compared to other countries like Italy and the USA. It reflects the influence of the broad spectrum of social distancing measures put in use by the government of India, which has played the role of a barrier to growing infected cases and deaths, apparently helped to slow down the epidemic growth. Our short-term forecast reveals that at the regional level, Delhi, Rajasthan, Gujarat, Maharashtra, Uttar Pradesh, Madhya Pradesh, Telangana, and Tamil Nadu will be the most affected states in the coming days. Considering the situation, lockdown should not be lifted in these states. The number of cases in Kerala and Karnataka is found to be reducing. Moreover, these states are shifted from the red zone to blue. Since very little growth in the future is predicted, lockdown may be lifted in these states with some restrictions for the proper functioning of economic activities. While states in green and light green zones, namely, Himachal Pradesh, Goa, Uttrakhand, Bihar, Jharkhand, Chhattisgarh, Odisha, Sikkim, Assam, Arunachal Pradesh, Nagaland, Manipur, Mizoram, Tripura, Meghalaya show very less growth in the infected cases till 1 May, therefore, lockdown may be uplifted there. On India level, there will be around 36335.63 [95% PI(30884.56, 42918.87)] cases and 1099.38 [95% PI(959.77, 1553.76)] deaths up to 1 May 2020. The forecasts presented here are based on the assumption that current mitigation efforts will continue.

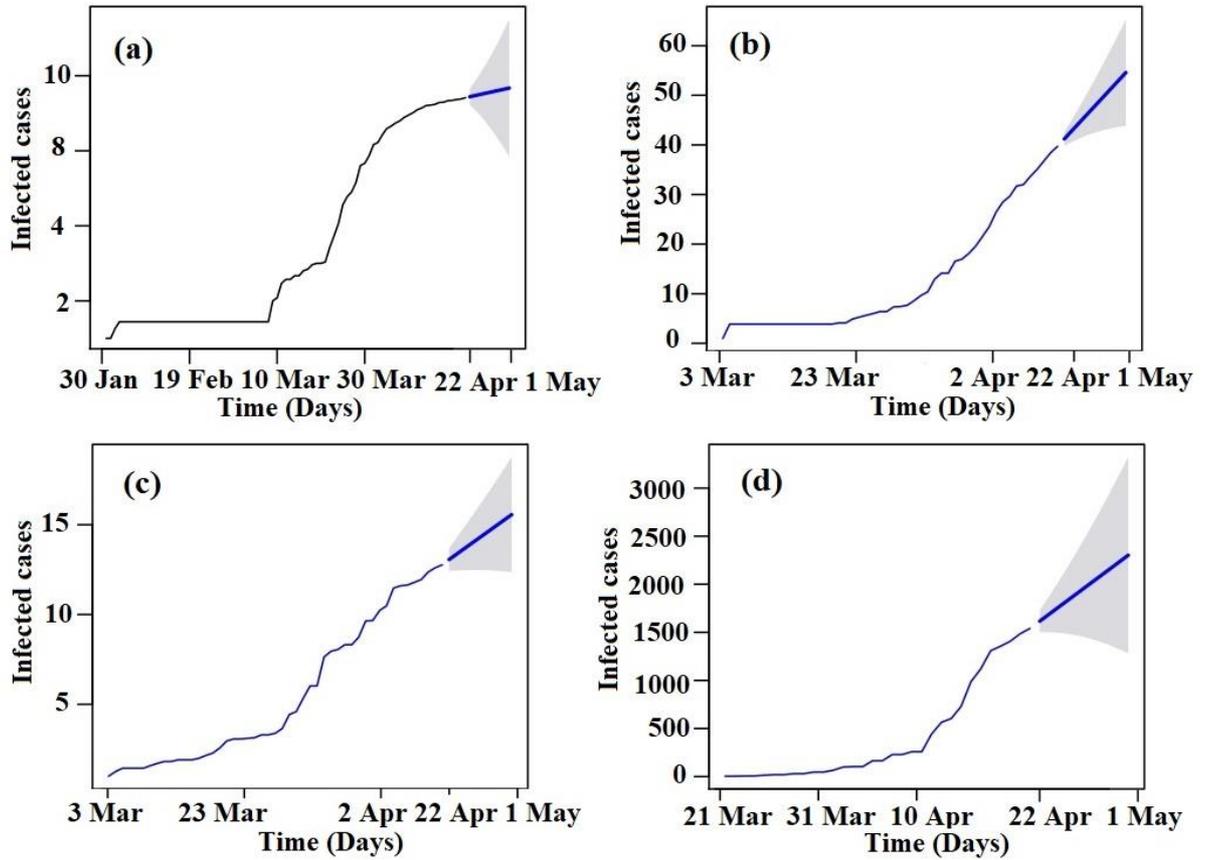

**Figure 5:** (a) 10-days ahead forecast (22 April to 1 May 2020) using Holt's Method for Kerala; (b) 10-days ahead forecast (22 April to 1 May 2020) using Holt's Method for Rajasthan; (c) 10-days ahead forecast (22 April to 1 May 2020) using Holt's Method for Delhi; (d) 10-days ahead forecast (22 April to 1 May 2020) using Holt's Method for Madhya Pradesh.

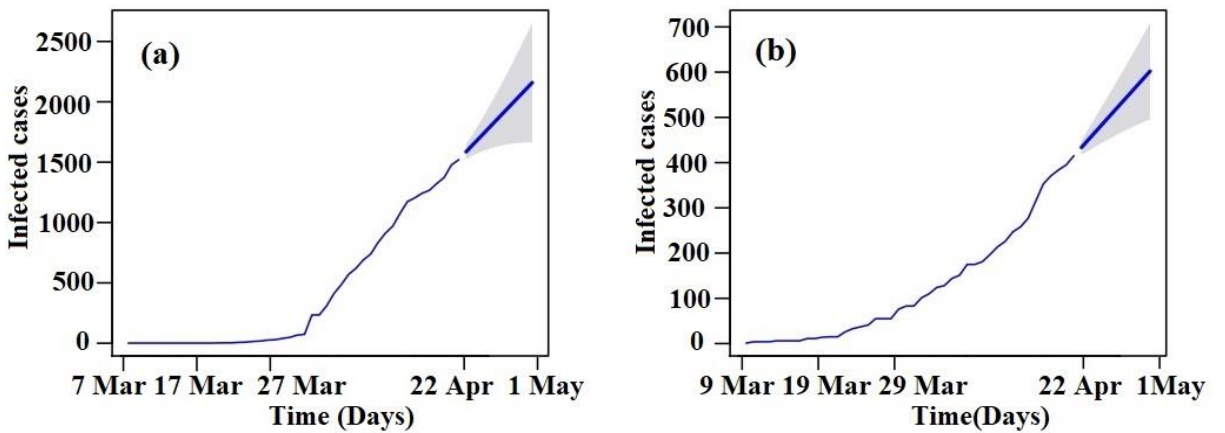

**Figure 6:** (a) 10-days ahead forecast (22 April to 1 May 2020) using Holt's Method for Tamil Nadu;(b) 10-days ahead forecast (22 April to 1May 2020) using Holt's Method for Karnataka.

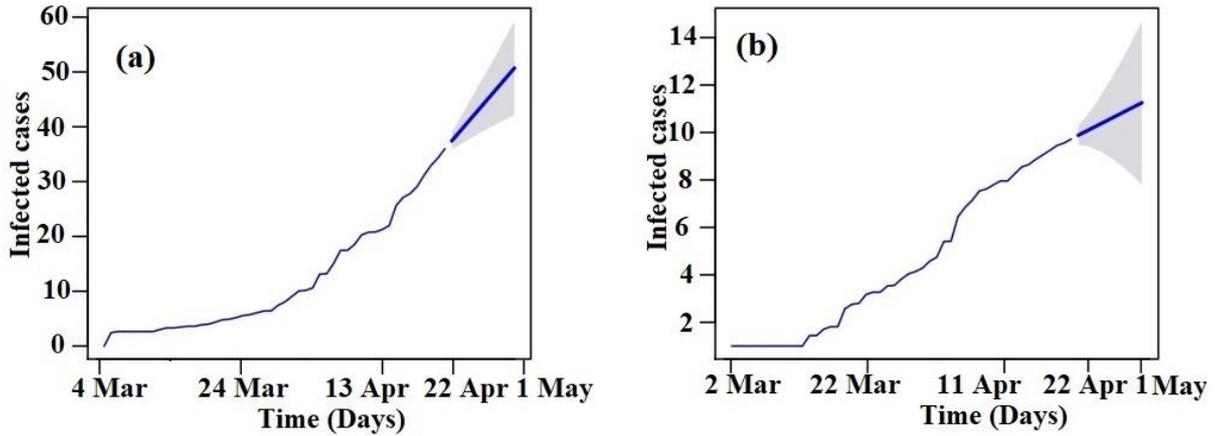

**Figure 7:** (a)10-days ahead forecast (22 April to 1 May 2020) using Holt's Method for Uttar Pradesh; (b) 10-days ahead forecast (22 April to 1 May 2020) using Holt's Method for Telangana.

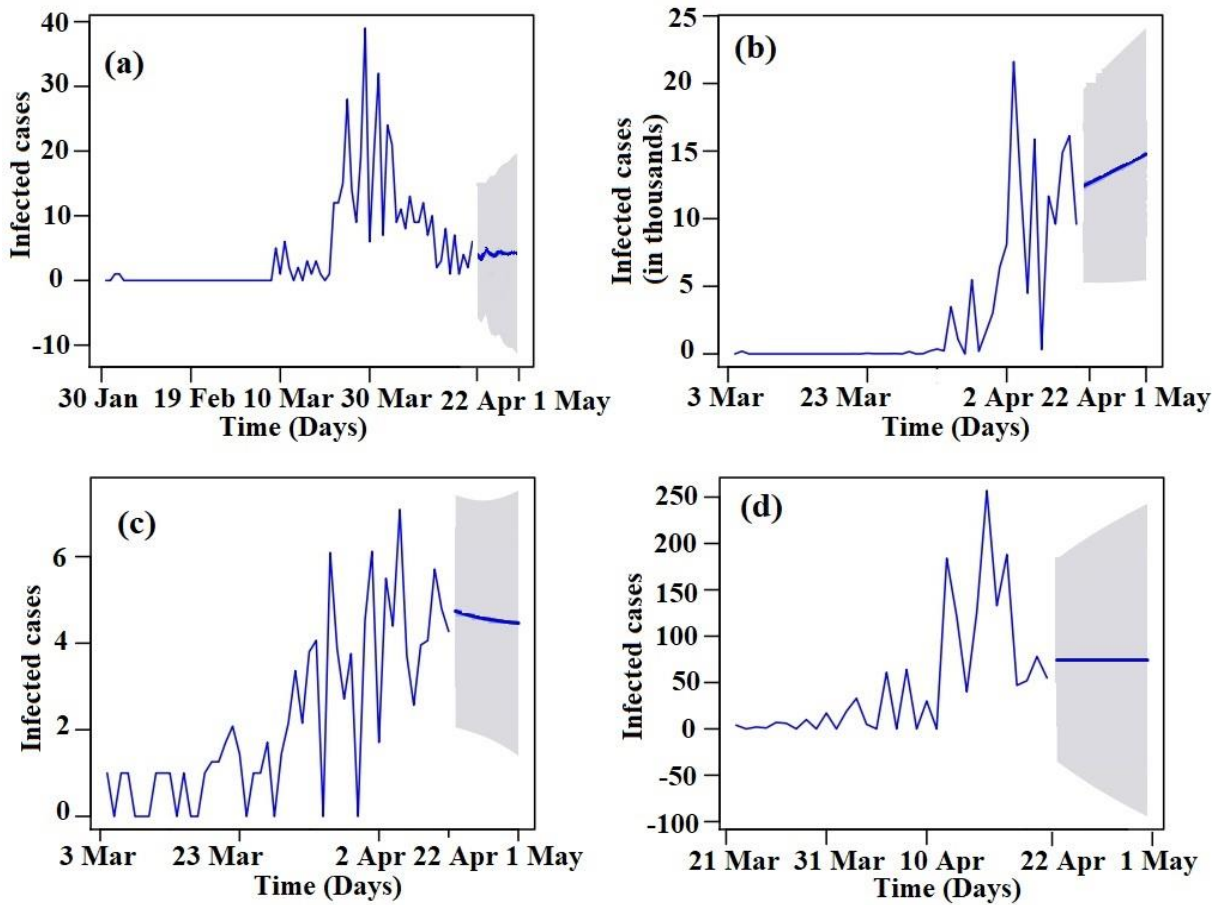

**Figure 8:** (a) 10-days ahead forecast (22 April to 1 May 2020) using ARIMA(2,1,0) model for Kerala; (b) 10-days ahead forecast (22 April to 1 May 2020) using ARIMA(0,1,1) model for Rajasthan; (c) 10-days ahead forecast (22 April to 1 May 2020) using ARIMA(1,1,2) model for Delhi; (d) 10-days ahead forecast (22 April to 1 May 2020) using ARIMA(0,1,1) model for Madhya Pradesh.

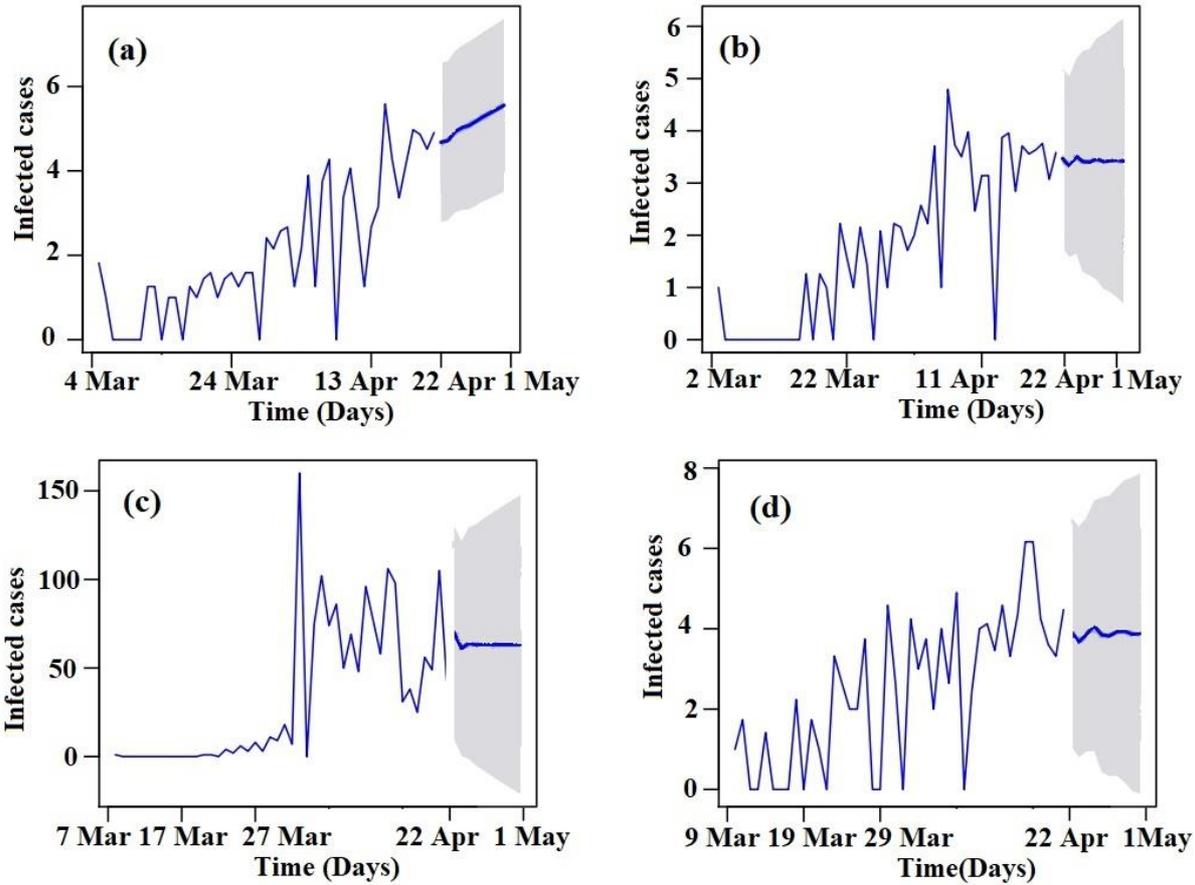

**Figure 9:** (a) 10-days ahead forecast (22 April to 1 May 2020) using ARIMA(2,1,1) model for Uttar Pradesh; (b) 10-days ahead forecast (22 April to 1 May 2020) using ARIMA(2,1,0) model for Telangana; (c) 10-days ahead forecast (22 April to 1 May 2020) using ARIMA(1,1,1) model for Tamil Nadu; (d) 10-days ahead forecast (22 April to 1 May 2020) using ARIMA(3,1,0) model for Karnataka.

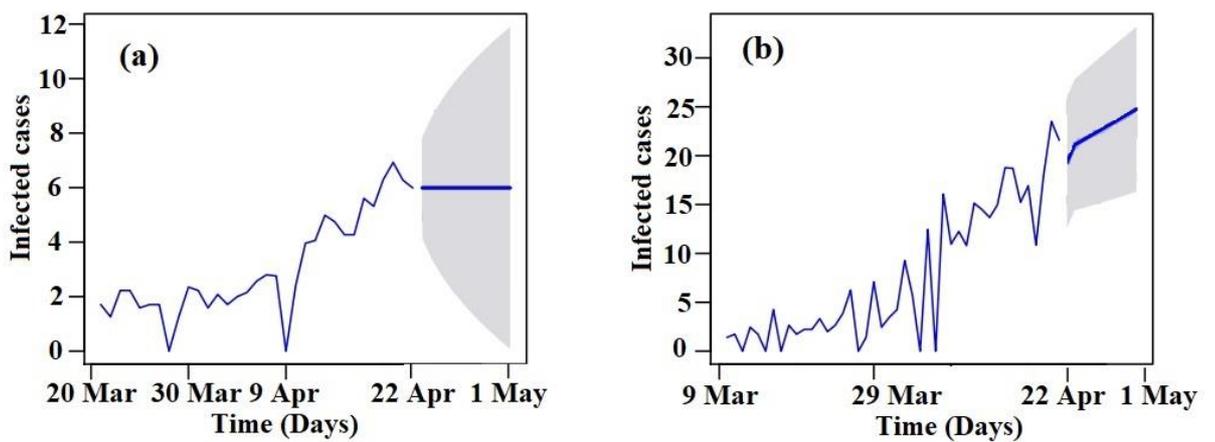

**Figure 10:** (a) 10-days ahead forecast (22 April to 1 May 2020) using ARIMA(0,1,0) model for Gujarat; (b) 10-days ahead forecast (22 April to 1 May 2020) using ARIMA(0,1,2) model for Maharashtra.

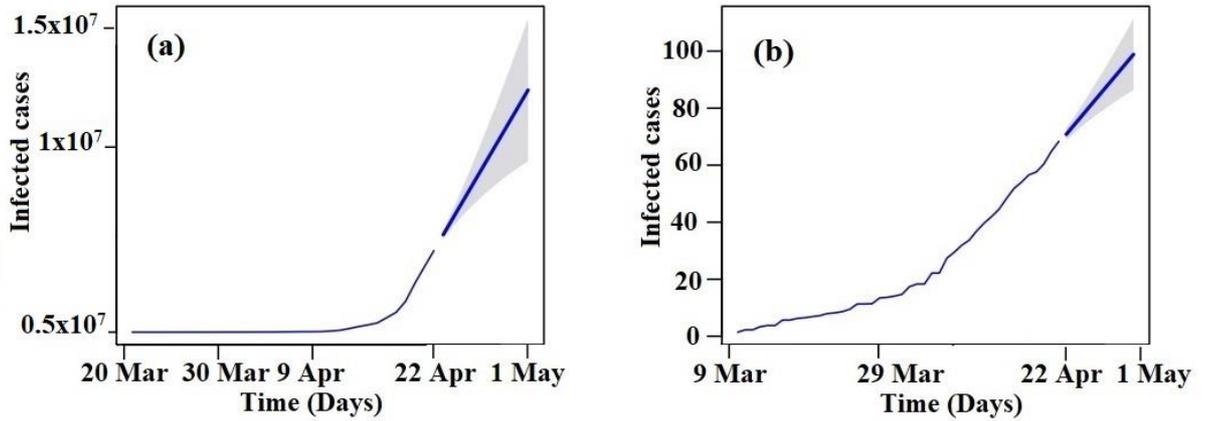

**Figure 11:** (a) 10-days ahead forecast (22 April to 1 May 2020) using Holt's Method for Gujarat; (b) 10-days ahead forecast (22 April to 1 May 2020) using Holt's Method for Maharashtra.

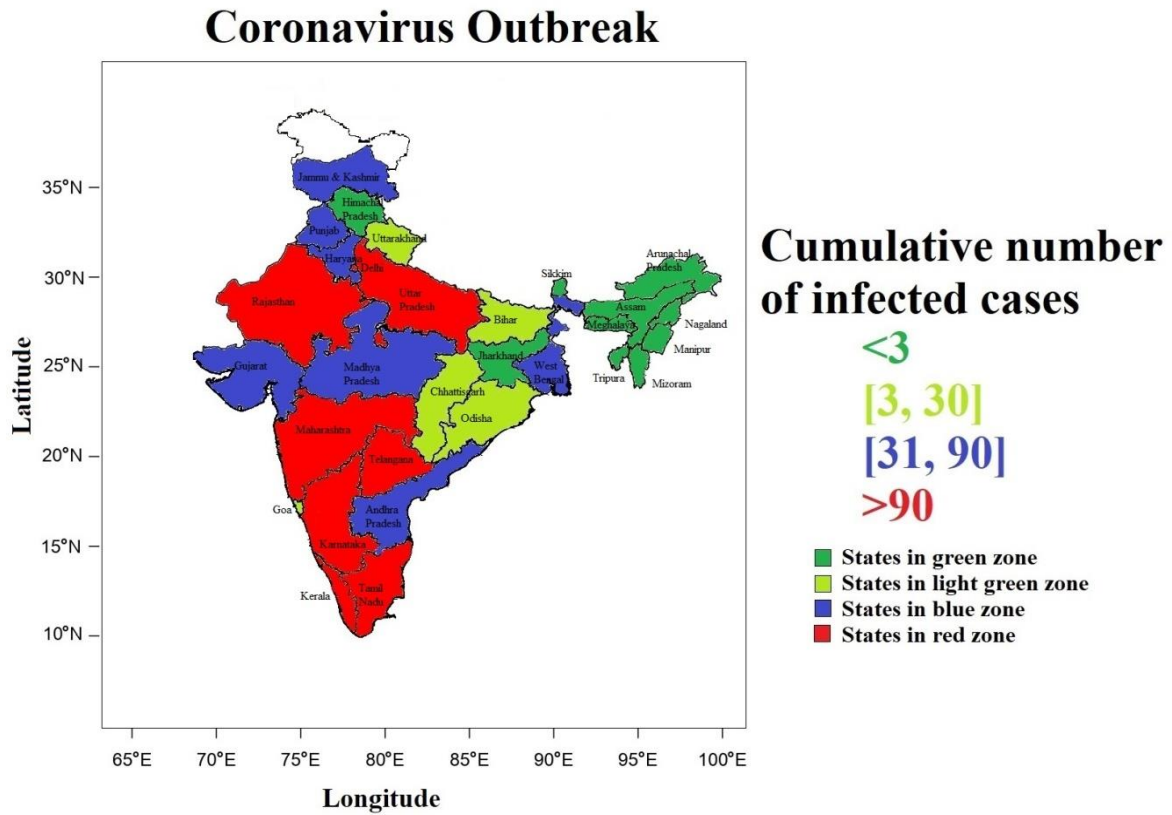

**Figure 12:** Spatial distribution of the coronavirus outbreak in the period of 30 Jan to 1 April 2020.

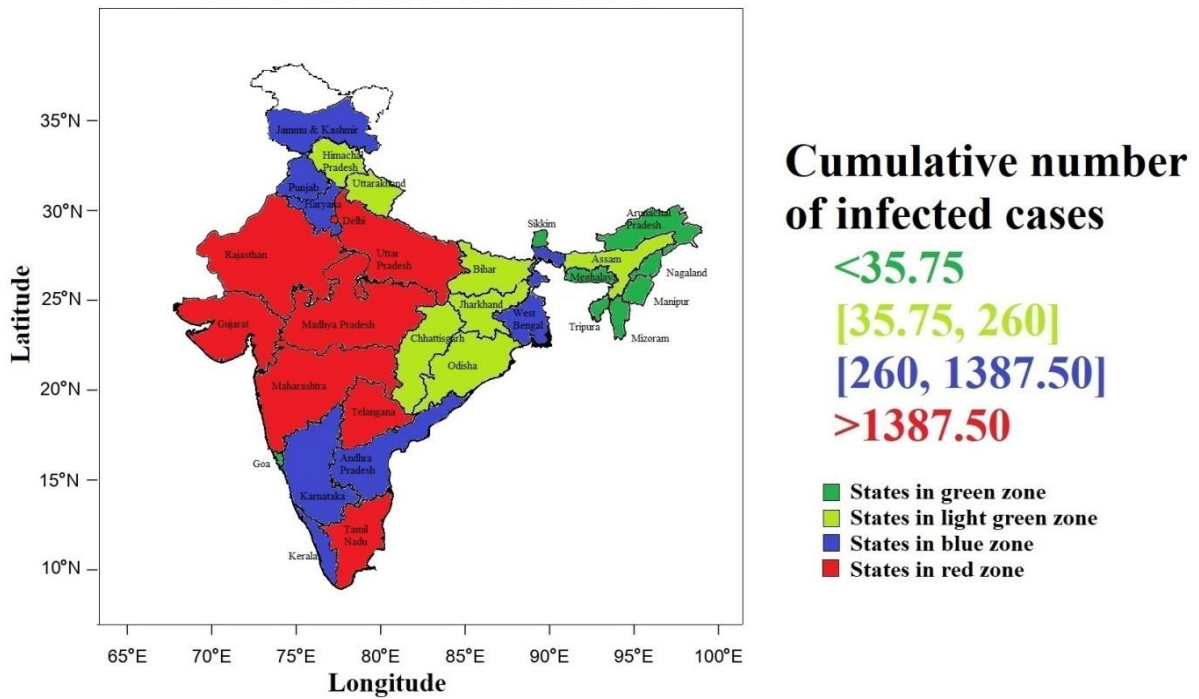

**Figure 13:** Spatial distribution of the coronavirus outbreak in the period of 30 Jan to 1 May 2020.

## 5 Data Availability

We obtained daily updates of the cumulative number of infected cases and deaths of the corona-virus illness for India from Worldometer website (online available: **https://www.worldometers.info/corona-virus/country/india/**). To obtain the state-wise cumulative number of infected cases and deaths for the corona-virus illness we used the government of India website (online available: **https://www.mygov.in/corona-data/covid19-statewise-status**). We gathered data of infected case(s) every day at 12 midnight (GMT-5) from 30 January to 21 April 2020. And forecasted the cumulative number of infected cases and deaths of the epidemic over the India and the cumulative number of infected cases in ten Indian states: Kerala, Maharashtra, Delhi, Gujarat, Tamil Nadu, Telangana, Uttar Pradesh, Madhya Pradesh, Karnataka, and Rajasthan, which show a high burden of COVID-19 cases.

## 6 Conflicts of Interest
The authors declare no conflicts of interest.

## 7 FundingStatement
Research Support is provided by the Indian Institute of Technology Mandi.